\documentclass[10pt]{article}
\usepackage[english]{babel}
\begin{document}
\begin{center}
{\bf \huge
TOPOLOGY AND BEHAVIOUR OF AGENTS: CAPITAL MARKETS
}
\end{center}

\begin{center}
Ondrej Hudak, \footnote{\small present e-mail: hudako@mail.pvt.sk}
\end{center}

\begin{center}
Faculty of Finance, Matej Bel University, \\ Cesta na amfiteater 1, SK-974 15 Banska Bystrica, Slovak Republic,\\
e-mail: ondrej.hudak@umb.sk
\end{center}

\begin{center}
Jana Tothova
\end{center}

\begin{center}
Stierova 23, Kosice, SK-040 23 Slovakia
\end{center}

\newpage
\section*{Abstract}
On a capital market the social group is formed from traders. Individual behaviour of agents is influenced  by the need to associate with other agents and to obtain the approval of other agents in the group. Making decisions an individual equates own needs with those of the other agents from the group. Any two agents from the group may interact. The interaction consists of the exchange of information and it costs some money. We assume that agents interact in such a way that they give reference to the origin of the information if asked by other agents. Thus the agent may verify obtained private information. Hudak recently  used methods described by Rivier to study social behaviour of such agents. He characterized the quantity which corresponds to verification of information. Here we study a capital market and social behaviour of agents relations. Quantity which characterizes verification of information contributes to an aversion of an agent with respect to a risk. The mix of investments of an agent in a given cell with an average measure A of risk aversion in the cell is found from minimum of the average  per cell aim function $<FM>$. Absolute minimum corresponds to such a state in which there is an optimal mix of the exchange of information for a given expectations about the capital market. The crowd and personal /$\approx <f>$/ contributions to the risk aversion of an agent are present in the aversion constant A.  We have discussed a stable and metastable states of the market for different values of E, an expected return for a given investment period, of EV, an expected risk for a given investment period, and of b, a constant which characterizes contribution of the quantity $<f>$ to the risk aversion. Characteristics of the variance of n for the distribution of nonreducibile subgroups are found. Our model describes intermediary process effects.

\newpage
\section{Introduction}

In a social group its members are called agents.
The social group is characterized in general in accordance with
\cite{1} - \cite{5}: we assume that individual behaviour
of agents is influenced to some degree by the need to associate
with other agents and to obtain the approval of other agents in
the group which is characterized by a very large nonrational
and emotional element to decisions of agents. Making
decisions an individual equates own needs with those
of the other agents from the group. We assume that any two agents from the group
may interact. The interaction consists of the exchange of information
and it costs some "energy" e.i. there are costs associated with the exchange of information. 
We assume that the information is well defined, and we
assume that agents interact in such a way that they give reference
to the origin of the information if asked by other agents.
The information is private \cite{6}. Public information is not studied in this paper.
Thus there is present information asymmetry.
We assume that the agent may verify obtained  information.
It means that the agent is giving references to two and only two (here we assume two and only two foe simplicity) with which he/she exchanged information. These two agents may be then called by the law court to give evidence about the information which they exchanged with the agent in a case of a law trial. Unverifiability is itself a special kind of information problem \cite{6}. Then there exits a subgroup of interacting agents the interaction
of which has the following property: it is nonreducible \cite{5}.

A cell is such a configuration of a number of nonreducible
subgroups in which every two interacting agents belong to two
nonreducibile subgroups /subgroups are connected in this sense/
and which is closed. Such a cell  may disappear and may be created,
may change number of nonreducible subgroups in a reversible way.
Because the structure, configuration of interactions between
agents in the group, forms a macroscopic structure we say that it is
a microreversibile process any process within a nonreducibile subgroup
and within a cell. Statistical equilibrium of the whole group
is characterized by a set of different subgroups of the type mentioned
above and by a probability that such a subgroup occurs. Thus we have
probability distribution which characterizes the group. Moreover
there exists an equation of state which enables to compare different
macroscopic states of the group. The statistical equilibrium due
to microreversibility is characterized by the maximum of entropy
and by the minimum of costs of information exchange minus a return.

We used \cite{5}
methods of statistical physics to study  social behaviour of agents,
mainly the presence of topological structure of interactions between
agents and its changes, which is the most important property of
the group of agents. There are three empirically observed dependencies of personal radius which enabled us
to characterize the quantities of cells,
faces /nonreducibile subgroups/, vertices /agents/ and bonds /interactions/ \cite{2}.
There exist constrains, such as a fixed number V of agents in
the group,  a number E of interactions within the group, a number F
of subgroups which are nonreducibile /faces/, and a number C of cells. Thus
we have a structure which is equivalent to random cellular networks.
Such networks and their evolution were described by Rivier,
\cite{3} and \cite{4}. Rivier applied methods
of statistical mechanics to study these networks.

Three empirically observed  dependencies of personal radius dependence on some factors
enabled us \cite{5} to characterize the quantity F which characterizes verification of information.
In this paper we study capital market and social behaviour of agents. Quantity F, and in general $<f>$ the average number of nonreducibile subgroups per cell, is the
quantity which characterizes verification of information and thus contributes to aversion of agent with respect to a risk.
The mix of investments of an agent in a given cell with an average measure A of risk aversion in the cell is found from
extremes of the function FM, the aim function. We find the minimum of the average  /per cell/ function $<FM>$. An absolute minimum of the function $<FM>$ corresponds to such a state 
in the group of agents in which there is an optimal mix of exchange of information for a given expectation of return and a given expectation of risk on the capital market.
Finding a minimum of the aim function $<FM>$ with respect to the parameter $ <n> $ of the group of agents we find the optimal mix of exchange of information. We assume that the number of nonreducible subgroups F per cell $<f>$ contributes linearly to aversion constant A, $ (A = A_{0} + (<f> - 4).b)$. The crowd /$A_{1} = A_{0} - 4.b $/ and personal /$\approx <f>$ / contributions to the risk aversion of an agent are thus present. We use $\frac{<f>.<n>}{3}$ as an average number of agents per cell.

We studied in this paper capital market and social behaviour of agents. We have formulated model describing behaviour of agents on the capital market.
We have found that a pyramidal structure appears in the group, $<f> = 4$, then $A = A_{0} $.
This type of structure  corresponds to hierarchical economy systems.
We have found that when the structure
contains topologically only one cell then $<n> = 6$ and A tends to infinity. Better verification of information leads to expectation of higher returns and thus the acceptable risk is larger.
This type of structure  corresponds to market-based economy systems.

In practice both market- and hierarchy- based systems co-exist in modern economy. There exist open markets and there exist large organizations.
We have found that this co-existence coresponds to a state which is the minimum of the aim function is in the range $ 3 < <n> <6$. In our model the market-based system coresponds to a state with $<n> = 6$ and the hierarchy-based szstem to a state with $<n> = 3$.

The mix of investments of an agent in a given cell with an average measure of risk aversion in the cell is found from
extremes of the function FM, the aim function. In our model we included intermediary property of firms - cells. Thus we have formulated a general model for topology of exchange of information and behaviour of agents on capital markets which may take into account intermediary property. We find the minimum of the average  /per cell/ function $<FM>$ for the simpler case in which there is zero intermediary. Thus assume that in this paper in our model the individual investors and agents from financial intermediary firms do not differ. More general case will be studied elsewier.

The extreme of the aim function $<FM>$ is minimum for some conditions on the market and in the group of agents. This minimum corresponds to $<n>$ between 3 to 6 as we mentioned above.
This is a stable state. The extreme may be also a maximum. Then there may be two minima of the function FM or one minimum. One of them is that which corresponds to $<n> = 3$ /the pyramidal hierarchical structure/ which is either a metastable state either a stable state,
depending on conditions on the market and in the group of agents.
 The other one corresponds to $<n> = 6$  which is a stable state. We call efficient capital markets markets which are of the second type. We have characterized variance of n for the distribution of nonreducibile subgroups. This variance diverges to infinity with $<n>$ tending to 6. Thus the market-based economy system with $ <n> $ near 
   or equal to has very large variance of nonreducibile subgroup number of agents.

\section{Capital Market and Social Behaviour of Agents}

We know \cite{5}  that a personal diameter r increases when F decreases. We generalize this relation to every cell: we assume that personal diameter r increases when $<f>$, the average number of nonreducubile groups per cell, decreases, $<f>$ is an average number of faces per cell in language of topology. Quantity F, and generalizing quantity $<f>$, is the quantity which characterizes verification of information. The larger F, and generalizing $ <f> $, the more verified information in general, e.i. not only a specific information exchanged by members of a subgroup. Aversion of an agent to risk on capital
markets also contains characterization of verification of information. We assume that the number of nonreducible subgroups F per cell, $<f>$, contributes linearly to this aversion constant A:

\begin{equation}
\label{1}
A = A_{0} - 4.b + <f>.b .
\end{equation}

Here $ A_{0}$ is a risk aversion constant for $<f> = 4$ when $A = A_{0} $, b is a constant which characterizes contribution
of the quantity $<f>$ to the risk aversion. The larger quantity $<f>$ the better verification of information, the larger expected return contribution of the investment and thus the larger acceptable risk.

Note that when the pyramidal structure appears in the group, $<f> = 4$, then $A = A_{0} $.
This type of structure  corresponds to hierarchical economy systems, for these systems see in \cite{6}. The aversion constant A has minimum value as concerning dependence on the quantity $ <f> $ in this case. Then lower risk is acceptable and correspondingly lower profit is acceptable also, especially for low $ A_{0} $ constant. The constant A does not contain contribution from the verification of information. Due to very low risk inthis case price signals do not work in these systems. Information on resources and aims and objectives, is flowing /exchanging/ through the hierarchy to the decision makers.

Note that when the structure
contains topologically only one cell then $<n> = 6$ and A tends to infinity. Better verification of information leads to expectation of higher returns and thus the acceptable risk is larger.
This type of structure  corresponds to market-based economy systems, for these systems see in \cite{6}. They work via price signals.
We interpret $(<f>.b)$ as an average measure of risk aversion of a personal /individual/ contribution to risk aversion different from the crowd contribution of this person $ A_{0} - 4.b $ in the cell, because F and $<f>$ are characteristics of verification of information. Trader - agent supplies or demands goods and services /on the capital market shares or bonds/ if the market price exceeds or undervalues his/her own valuation of shares or bonds. To make own valuation of shares or bonds the agent individually verifies information on it.

In practice, as noted in \cite{6}, both types of system involve interaction of people: in the market system they interact  as traders, in the hierarchy as agents within an organization. Both market- and hierarchy- based systems co-exist in modern economy \cite{6}: we have open markets and we also have large organizations such as the joint-stock company.

\section{A financial intermediary}

In market economies the key role of providing financial intermediaries and transaction services play banks and other financial firms. Intermediation process is understood well in a hypothetical market economy by considering how it might function without financial intermediaries \cite{7}. For this economy there is assumed for simplicity that government receipts equal outlays and that the exports of goods and services equal import. It is also assumed that all savings is by households and all investment is by businesses. Household claim on business may be in the form of debt or equity capital. This is minor additional difference from reality \cite{7}. The level of interest rates and the associated valuation of equity in this economy is determined by supply and demand for savings, the influence of any monetary authority is not taken into account \cite{7}. In the absence of intermediaries households would have to hold their savings in the form of equity or debt claims on specific firms.
They are limited in their ability to diversify their holdings by the high costs of obtaining information about many different companies and of dealing with a large volume of small denomination securities, \cite{7}.  Thus each household would have to expect the large risk due to particular firms in which it invested. This could result in  a loss of most or all of its savings. Such a large risk would lead to a high rate of interest and cost of capital which a household would have to take. Reduction of their risk would lead to households willing to accept a lower rate of interest. The financial intermediaries play a key role, \cite{7}, to reduce the risks faced by households /individual savers/, by pooling their savings and using these to assemble diversified portfolios of assets. Diversification is such that risks of bankruptcy of different firms do not depend on the same economic conditions specific to the firm, industry, geographic region or the entire economy. This diversification requires specialized knowledge and expertise, and a large portfolio. Low, or even negative, correlation among the returns on different assets in portfolio and larger number of these assets reduce risk. Thus intermediaries can pay a lower rate of interest to households. Competition among financial intermediaries leads together with lower rate of interest payed to households to lower costs of financing, which leads to lower costs of financing firm enterprises, \cite{7}.
This hypothetical market economy we will use in this paper to describe intermediary process quantitatively.

In our model of a capital market every point /vertex/ corresponds to an agent acting on the market. In the simplified model economy described above these agents would be households. Financial intermediaries which lead to lower costs will be described in our model as cells with a given risk aversion constant of the firm to which risk aversion constants of agents from this firm - cell should not be too far. Some other cells are cells with zero risk tolerance cell constant, agents in these cells corespond to households /individual savers - agents/. Thus we have in our model not only households but also financial intermediaries description included.

\section{The mix of investment and interaction of agents}

We will assume that every agent has its objective to maximize its expected
utility of wealth \cite{8}. If returns are normally distributed and the investor has constant absolute risk aversion, then expected utility can be written as \cite{9}:

\begin{equation}
\label{2'}
EU = - exp{(-c(E-\frac{c}{2}.EV))}
\end{equation}

where the risk tolerance constant is $c = \frac{1}{A}$, A is the risk aversion constant, E is the expected value of end-of-period wealth, EV is the expected variance of end-of-period wealth. To make this utility as large as possible, one maximizes \cite{10}:

\begin{equation}
\label{2''}
E - \frac{c}{2} . EV.
\end{equation}

This leads to maximum expected utility of an agent, assuming that the risk aversion is constant.
Agents are interacting, some of them correspond to households /individual investors/, some of them are from the financial intermediary firms. In our model we will assume that every agent has the same E / the expected value of end-of-period wealth, in percents/ and the same EV /the expected variance of end-of-period wealth, in percents/. Thus c is a number, and A is also a number. In reality our assumption is not true, there exist dispersion of E and EV quantities. Assuming the same E and EV quantities we assume that there exist a mean value of E and Ev for the dispersion.

We would like to consider not only maximization of expected utility of an agent, but also optimization of interactions of agents in such a way in which interacting agents corresponding to households and to intermediary financial firms maximize their expected utility.
While the first case corresponds to maximization of (\ref{2''}), the second case is more general.
To maximize the expected utility of an agent in fact it is necessary to maximize the function:

\begin{equation}
\label{2'1}
fM = c(E-\frac{c}{2}.EV)
\end{equation}

In this function there is a constant c. This constant is dependent on the constant A which depends on the quantity $<f>$. Optimization of interactions of agents in such a way in which interacting agents corresponding to households and to intermediary financial firms maximize their expected utility leads to optimization  of (\ref{2'1}) for every agent taking into account the structure of a cell. The cell may represent agents - households, but also a financial intermediary.
The first one cell will be described in our model as a cell with different aversion constants of agents from this cell, the second one will be described in our model as a cell with almost the same aversion constant of agents from this cell.
To model this fact we will optimize the following function:

\begin{equation}
\label{2'11}
fM = \sum_{i=1}^{i=V_{C}} c_{i}(E-\frac{c_{i}}{2}.EV) - \sum_{i=1}^{i=V_{C}} \gamma (c_{i}-c_{B})^{2}
\end{equation}

Here $V_{C}$ is a number of agents in the cell C.
The first sum is a sum of terms of the type (\ref{2''}), the second term is a sum which describes how far are constants $c_{i}$ of agents from a constant $ c_{B} $ which is characteristic for a given cell. The constant $\gamma$ is a positive or zero constant. If it is zero, then the function fM describes a sum of functions fM for single agents. This correspond to individual investors in the sense that their aversion to risk constant is different, individual. If it is nonzero, then the function fM describes a sum of functions fM for agents which are not individual investors in the sense that their aversion to risk constant is not too much different. These agents are agents from an intermediary financial firm characterized by a constant $ c_{B} $. If the constant $\gamma$ is very large, then all the constants $c_{i}$ from this cell tend to the same value $ c_{B} $ characteristic for the firm.

To interpret the second sum in terms of expected returns and expected variances let us rewrite the function fM (\ref{2'11}) in the form:

\begin{equation}
\label{2'111}
fM = \sum_{i=1}^{i=V_{C}} (c_{i}((E + 2.c_{B}.\gamma)-\frac{c_{i}}{2}.(EV + 2.\gamma)) - c_{B}^{2}. \gamma )
\end{equation}

We see that the second term in (\ref{2'11}) corresponds to higher total expected value $E + 2.c_{B}.\gamma$ in which besides an expected value E of the individual investor a new contribution $2.c_{B}.\gamma$ appears. This contribution is due to the firm financial intermediary, due to diversification which is such that risks of bankruptcy of different businesses do not depend on the same economic conditions specific to the firm, industry, geographic region or the entire economy.
We see that the expected return of an individual investor should be higher to have the same expected value as total expected return /we are taking always a return for a given period/ for the financial intermediary firm. This diversification requires, as written above, specialized knowledge and expertise, and a large portfolio. The higher risk tolerance constant $ c_{B} $ of the firm leads to the higher total expected return.

We see further that the second term in (\ref{2'11}) corresponds to higher total expected risk $EV + 2.\gamma$, in which besides an expected variance EV of the individual investor a new contribution $2.\gamma$ appears. This contribution is again due to firm financial intermediary, due to
diversification which is such that risks of bankruptcy of different businesses do not depend on the same economic conditions specific to the firm, industry, geographic region or the entire economy. This diversification requires, as written above, specialized knowledge and expertise, and a large portfolio thus larger expected risk may be accepted.
We see that the expected risk of an individual investor should be higher to have the same  expected risk as total expected risk for the financial intermediary firm.
The stronger tendency /the higher value of the constant $\gamma$ / in the firm to have the same risk tolerance constant $ c_{B} $ for all agents in the firm the higher total expected variance - risk. The constant contribution $- c_{B}^{2}. \gamma$ gives a term in fM which is proportional to the number of agents in the cell - firm. This contribution is larger for larger risk tolerance constant $ c_{B} $ /in absolute value/ and is negative.

Optimization of (\ref{2'111}) for given E, EV and $ c_{B} $ leads to optimization of the number of agents in the cell and to optimization of verification of information which is characterized by f. Correspondingly optimization for the whole group /agents on capital market/ for given E, EV and $ c_{B} $ leads to optimization of the number of agents in the cells and to optimization of verification of information on the market, which is characterized by $<f>$.
Then structure of exchange of information in the group of agents will be such that we obtain the highest value of the function fM.

We see further that the second term in (\ref{2'11}) for $\gamma$ very large leads to all risk tolerance constants in a cell to be the same.
We see also that the second term in (\ref{2'11}) for $\gamma = 0$, e.i. for vanishing $\gamma$,
leads to all risk tolerance constants in a cell to be different.

To obtain an aim function FM for such a cell, we will consider function fM from (\ref{2'11}) with negative sign.

The mix of investment of a single cell in the mean field approximation is found from the function:

\begin{equation}
\label{2}
<FM> = (- c.E_{B} + \frac{c^{2}}{2}EV_{B}) \frac{<f>.<n>}{3} + \gamma . c_{B}^{2}. \frac{<f>.<n>}{3},
\end{equation}

where $E_{B} = E + 2 \gamma c_{B}$ is an expected return for a given investment period modified by a contribution of the return from the cell /firm/ financial intermediary, $EV_{B} = EV + 2 \gamma $ is an expected risk for a given investment period modified by contribution of the risk from the cell /firm/ financial intermediary, E and EV is an expected return and an expected risk respectively for a given investment period. Here $c_{B}$ is a risk tolerance constant given for the cell.

The mix of investment of a given cell /financial intermediary firm/ with an average measure c of risk tolerance of a person /which is inversely equal to a crowd contribution of a person plus a personal /individual/ contribution of this person in the cell/ is found from the average  /per cell/ function $<FM>$ above (\ref{2}). Let us now study the case $\gamma = 0$, which does not distinguish between individual investors /for which $\gamma = 0$ / and a financial intermediary firm.

\section{The case in which individual investors and a financial intermediary firm are not distinguished}

The case in which individual investors and a financial intermediary firm are not distinguished we will study now. Thus we study the case in which financial intermediary in the capital market is not considered, all agents of such financial intermediary firm behave as individual investors. The case $\gamma > 0$, which does distinguish between individual investors /for which $\gamma = 0$ / and a financial intermediary firm / for which $\gamma > 0$ / will be studied elsewhere.

Thus the equation for the aim function FM is found from (\ref{2}):

\begin{equation}
\label{3}
<FM> = (- c.E + \frac{c^{2}}{2}EV)\frac{<f>.<n>}{3},
\end{equation}

where $\frac{<f>.<n>}{3}$ is an average number of agents per cell. From (\ref{1}) and (\ref{3}) it has the form, note that $c = \frac{1}{A} = \frac{1}{A_{0} - 4.b + <f>.b}$:

\begin{equation}
\label{4}
<FM> = (- \frac{E}{A_{0} - 4.b + <f>.b} + \frac{EV}{2.(A_{0} - 4.b + <f>.b)^{2}})\frac{<f>.<n>}{3}
\end{equation}

and where for an equilibrium structure with a given number of cell C, of faces F, of interactions E, and of agents V, see \cite{3} and \cite{4}:

\begin{equation}
\label{5}
<f>= \frac{12}{(6-<n>)},
\end{equation}

and where $A_{1} = A_{0} - 4.b$. From the equation for the aim function FM (\ref{4}) we will now find which number of cell C, of faces F and of interactions E it minimizes, here number of agents V is given. It means that there is a minimization of FM with respect to $<n>$, number of interaction in an ireducibile subgroup /face/. Thus we are looking for such a structure of agents on the capital market which gives the lowest value of the aim function FM taking into account their interaction exchanging information. The case which does distinguish between individual investors and a financial intermediary firm is more complicated and will be studied elsewhere.

\section{Extremes of the function FM}

The function FM has extremes. Absolute minimum corresponds to such a state in which there is an optimal mix. There exists an extreme of FM $<n>'$ which is given by:

\begin{equation}
\label{6}
<n>'= 6.\frac{(EV (A_{0}-2b)- 2E(A_{0}-2b)^{2})}{( EV.A_{0}-2E(A_{0}-4b)(A_{0}-2b))}.
\end{equation}

Note that when contribution of nonreducibile subgroups is absent $b=0$, then $<n>'=6$. Thus only
one cell exists with many nonreducibile subgroups in this case.
Note that when the crowd and personal contributions to the risk aversion of an agent are present then $b>0$, $<n>' \neq 6$ and more than one cell maz exist in the group. There are several subgroups -cells. Note that
when the crowd risk aversion is zero, $A_{0} = 4.b $, then:

\begin{equation}
\label{6.1}
<n>'= -3(1 + \frac{4bE}{EV})
\end{equation}

which is less than 3, for E and EV positive. The quantity $ <n>' $ tends to 0 when $A_{0}$ tends to $2.b $. Many cells exist in the group - pyramidal structure may exist
for $ <n>' = 3 $. Thus it is necessary to discuss minima of the aim function
FM.

\section{The extreme is minimum}

The extreme may be an absolute minimum for $A_{0} < 2b$. This is the case when the tendency to verify information is stronger. The extreme (\ref{6}) is now minimum if:

\begin{equation}
\label{7}
2E(2b - A_{0})> EV > E \frac{A_{0}}{b} (2b - A_{0})
\end{equation}

and here the state with $<n> = 6$ has lower value of the aim function than the state $<n> = 3$.

The extreme is also minimum if:

\begin{equation}
\label{7.1}
E \frac{A_{0}}{b} (2b - A_{0}) > EV \geq 2E \frac{A_{0}(2b -A_{0})}{4b - A_{0}},
\end{equation}

and here the state with $<n> = 3$ has now lower value of the aim function than the state $<n> = 6$.

In both cases the extreme is a state in which some type of hierarchical structures exists locally in the group, the group is not hierarchical as whole.
The extreme - minimum gives the average number of agents per nonreducibie subgroup which is less than 6 and larger than 3.

The extreme (\ref{6}) is also minimum if:

\begin{equation}
\label{7.2}
EV > 2E(2b - A_{0}).
\end{equation}

Here however the state with $<n> = 6$ is minimum due to condition that $3  \leq <n> \leq  6$ is not fulfilled for $<n>'$, this extreme is larger than 6.
In this case the minimum of the aim function with $<n> = 6$ is a state in which there is no type of hierarchical structure locally present in the group.

The extreme (\ref{6}) is also minimum if 

\begin{equation}
\label{7.3}
0 \leq EV < E \frac{A_{0}}{b} (2b - A_{0}).
\end{equation}

Here the state with $ <n> = 3 $ is minimum due to the condition that $ <n> $ is from the interval $ 3 \leq <n> \leq $, the extreme has lower value than 3. In this case the aim function with $ <n> = 3 $ has minimum value in this interval. The state $ <n> = 3 $ corresponds to an hierarchical state.

\section{The extreme is maximum}

The extreme may be an absolute maximum for $A_{0} > 2b$. In this case the state $<n> = 6$ has always lower value of the aim function than the state with $<n> = 3$.

The extreme (\ref{6}) is maximum if, for $A_{0} > 4b$. :

\begin{equation}
\label{8}
EV > 2E \frac{A_{0}(A_{0}-2b)}{A_{0}-4b}
\end{equation}

and then there exist two states which minimize function FM /e.i. maximize return and minimize risk/.

The first one is with $<n> = 3$.
It gives the average number 3 of agents per nonreducibile subgroup, this number means that there is a pyramidal /hierarchical/ structure.

The second one is with $<n>=6$.
It gives the average number 6 of agents per nonreducibile subgroup, the value of the aim function FM is such that this state is much more stable than the state above /the state corresponding to $<n> = 3$/ which is a metastable state. The number $<n> = 6$ means that there is a one cell structure. Efficient capital markets are markets which are of the second type /strong stability, conservative and aggressive agents are
present on the market, return is high/ in our paper.

The extreme (\ref{6}) is maximum if, for $A_{0} > 4b$. :

\begin{equation}
\label{8.1}
2E \frac{A_{0}(A_{0}-2b)}{A_{0}-4b} > EV > 2E \frac{(A_{0}-4b)(A_{0}-2b)}{A_{0}}
\end{equation}

however then there exists one state which minimize function FM /e.i. maximize return and minimize risk/ for $3 \leq <n> \leq 6$. The state with $<n> = 3$, which was a metastable state is now not the metastable state. The minimum od the aim function FM is for $<n>=6$.
Efficient capital markets are markets which are of this type.

The extreme (\ref{6}) is maximum if, for $4b > A_{0} > 2b$. :

\begin{equation}
\label{8.2}
EV < 2E \frac{A_{0}(A_{0}-2b)}{A_{0}-4b}
\end{equation}

This inequality cannot be fulfilled because right hand side is negative for positive expected returns E and for positive expected risk EV.

The extreme (\ref{6}) is maximum if, for $4b > A_{0} > 2b$. :

\begin{equation}
\label{8.3}
EV > 2E \frac{A_{0}(A_{0}-2b)}{A_{0}-4b}
\end{equation}

This inequality is fulfilled always. Thus the state with $<n>=6$ is now a state with minimum aim function. There is no metastable state in this case.

\section{Aboav relation for the case of capital markets}

Aboav relation \cite{4} for the case of capital markets tells us:

\begin{equation}
\label{9}
n.m(f,n)= 5.f - 11 - K.(f - 1 - n),
\end{equation}

where K is a parameter of the group independent of f. Aboav's law describes how many, $/m(f,n)/$, agents are
present in average in a nonreducibile subgroup in a cell neighbouring to a cell with f nonreducibile subgroups with agent average number n.

Variance of n for the distribution of agents in nonreducibile subgroups of a cell with f nonreducibie subgroups is due to Weaire \cite{4}:

\begin{equation}
\label{10}
<(n - <n>)^{2}> = <n^{2}> + <n>^{2}
\end{equation}

This variance may be calculated from m(f,n) which is equal to n \cite{4} to Weaire and we obtain:

\begin{equation}
\label{11}
<n^{2}> = <n.m(f,n)> = 5.<f> - 11 - K.(<f> - 1 - <n>)
\end{equation}

\begin{center}
$= \frac{60}{6 - <n>} - 11 - K.(\frac{12}{6 - <n>} - 1 - <n>)$.
\end{center}

We see that for states in which $<n>$ tends to 6, the variance diverges as  $ \frac{12.(5-K)}{6 - <n>}$.
Note that $- 1 \leq K \leq 2$, \cite{3} and \cite{4}.

Note that structures /stable states/ in both examples on capital market are such that their entropy /informational/ is maximum /risk is minimum/ and their return is maximum minimizing the function FM.

\section{Conclusions}

We studied capital market and social behaviour of agents. Quantity F, and in general $<f>$, is the
quantity which characterizes verification of information and thus contributes to aversion of an agent with respect to a risk. We generalized this relation to every cell: we assumed that personal diameter r increases when $<f>$, the average number of nonreducubile groups per cell, decreases, $<f>$ is an average number of faces per cell. Thus we assumed that aversion of an agent to risk on capital markets also contains characterization of verification of information. We assumed that the number of nonreducible subgroups F per cell, $<f>$, contributes linearly to this aversion constant A.

When the pyramidal structure appears in the group, $<f> = 4$, then $A = A_{0} $.
This type of structure  corresponds to hierarchical economy systems. When the structure
contains topologically only one cell then $<n> = 6$ and A tends to infinity. Better verification of information leads to expectation of higher returns and thus the acceptable risk is larger.
This type of structure  corresponds to market-based economy systems. They work via price signals.
We interpret $A_{0}  -  4.b$ as a kind of measure of risk aversion of crowd and we interpret $(<f>.b)$
as an average measure of risk aversion of a personal /individual/ contribution to risk aversion different from the crowd contribution of this person in the cell.

In practice both types of system involve interaction of people: in the market system they interact  as traders, in the hierarchy as agents within an organization. Both market- and hierarchy- based systems co-exist in modern economy: there are open markets and there are  large organizations.
This corresponds to our state for which the aim function has its minimum for $ <n> $ from the interval $ 3 < <n> <6$. In our model the market-based system corresponds to a state with $<n> = 6$ and the hierarchy-based to $<n> = 3$, which are boundary cases of this interval for values of $<n>$.

The mix of investments of an agent in a given cell with an average measure A of risk aversion in the cell is found from
extremes of the function FM, the aim function. We find the minimum of the average  /per cell/ function $<FM>$. We assume that in our model the individual investors and agents from financial intermediary firms do not differ. More general case will be studied elsewhere.

When $<f> = 4$ then the pyramidal /hierarchical/ structure exists in which there are only nonreducibie subgroups with $n = 3$. Thus we consider this structure as a structure with a uniform risk aversion constant $A_{0}$. Then a
personal contributions to the risk aversion constant of an agent are those contributions which are corresponding
to structures with $<n>  > 3$. Thus there is a linear contribution to the risk aversion constant which is proportional to $<f> - 4$. This corresponds to nonuniform risk aversion constants in the group of agents. This contains a personal contribution to
the risk aversion constant.
We use $\frac{<f>.<n>}{3}$ as an average number of agents per cell.

The extreme of $<FM>$ is minimum for some conditions on the market, and in the group of agents, see above. This minimum corresponds to $<n>$ between 3 to 6. This is a stable state. The extreme may be a maximum. Then there are two minima of the function FM, one of them is that which corresponds to $<n> = 3$ /the pyramidal hierarchical structure/ which is a metastable state, and the other one corresponds to $<n> = 6$ /the structure with hexagons in average and with one cell/ which is a stable state. Efficient capital markets are markets which are of the second type /strong stability, conservative and aggressive agents are present on the market, returns are high/ in our paper. Note that under some conditions of agents and the expectations 
about the capital market the hierarchy economy system may be more stable than the market-based economy system.
 Aboav's law describes how many, $/m(f,n)/$, agents are present in average in nonreducibie subgroups in a cell neighbouring to a cell with f nonredducibile subgroups with agent average number n. This enables us to characterize variance of n for the distribution of nonreducibile subgroups. This variance diverges to infinity with $<n>$ tending to 6.

The financial intermediaries play a key role to reduce the risks faced by individual savers, by pooling their savings and using these to assemble diversified portfolios of assets. This diversification requires specialized knowledge and expertise, and a large portfolio. This will be studied using our model with nonzero $\gamma $ constant corresponding to a firm with a given tolerance risk constant.

\section*{Acknowledgment}

The paper represents a part of results of the VEGA project 1/0495/03.


\begin{thebibliography}{A}
\bibitem{1}
T. Plummer, The Psychology of Technical Analysis, Rev. Ed., Probus Pub.Comp., Chicago-Cambridge, 1993
\bibitem{2}
D. Lewis, The Secret Language of Success, Carroll and Graf Pub. Inc., USA, 1989
\bibitem{3}
N. Rivier, Journal de Physique, C9 N12 T46 (1985) 155
\bibitem{4}
N. Rivier, Physica, 23D (1986) 129
\bibitem{5}
O. Hudak, Topology and Social Behaviour of Agents,  http://arXiv.org/abs/cond-mat/0312723 , 2003
\bibitem{6}
I. Molho, The Economics of Information, Blackwell Pub., Oxford, Malden, 1997
\bibitem{7}
A. Grenspan, Commercial Banks and the Central Bank in a Market Economy, in Readings on Financial Institutions and Markets, P.S. Rose editor, 5th ed., R.D. Irwin Inc., Homewood, Boston, 1993, p. 294
\bibitem{8}
J. von Neumann and O. Morgenstern, Theory of Games and Economic Behaviour, 3rd. ed., Princeton University Press, Princeton, 1953
\bibitem{9}
J. Lintner, The Market Price of Risk, Size of Market and Investor Risk Aversion, Journal of Business, April (1968)
\bibitem{10}
W.F. Sharpe, Integrated Aset Allocation, Financial Analysts Journal, September -October (1987)
\end{thebibliography}
\end{document}